\documentclass[pra,aps,reprint,superscriptaddress,twocolumn]{revtex4-2}
\pdfoutput=1
\usepackage{hyperref}
\usepackage{comment}
\usepackage{subcaption}
\usepackage{tikz}
\usepackage{lipsum}
\usepackage{graphicx,color}
\usepackage{amsmath}
\usepackage{bm}
\usepackage{amssymb}
\usepackage{caption}
\usepackage{pifont} 

\usepackage[all]{hypcap}
\usepackage{mathtools}
\usepackage{physics}
\usepackage{enumitem}
\usepackage{framed,enumitem} 
\usepackage{adjustbox}
\usepackage{threeparttable}
\usepackage{float}
\makeatletter
\let\newfloat\newfloat@ltx
\makeatother
\usepackage{algorithm}
\usepackage{algpseudocode}

\usepackage{ulem}
\usepackage{xcolor}

\begin{document}

\makeatletter
\long\def\@makecaption#1#2{%
  \par\vskip\abovecaptionskip
  \begingroup
   \small\rmfamily
   \leftskip\z@ \rightskip\z@ \parfillskip\@flushglue 
   \@make@capt@title{#1}{#2}\par
  \endgroup
  \vskip\belowcaptionskip}
\makeatother

	\title{Uncovering Latent Structures in Robust Pulse Sequences: A Model-Based Reinforcement Learning Approach for Adaptable Quantum Control}
	\author{Tobias Kiermeyer}
	\email{t.kiermeyer@tum.de}
	\affiliation{\footnotesize Technical University of Munich, TUM School of Natural Sciences, Department of Chemistry, Lichtenbergstra{\ss}e 4, 85748 Garching, Germany}
	\affiliation{\footnotesize Munich Centre for Quantum Science and Technology (MCQST), Schellingstra{\ss}e 4, 80799 M{\"u}nchen, Germany}
	\author{Thomas Heydenreich}
    \affiliation{\footnotesize Technical University of Munich, TUM School of Natural Sciences, Department of Chemistry, Lichtenbergstra{\ss}e 4, 85748 Garching, Germany}
	\author{L\'{e}o Van Damme}
    \affiliation{\footnotesize Technical University of Munich, TUM School of Natural Sciences, Department of Chemistry, Lichtenbergstra{\ss}e 4, 85748 Garching, Germany}
    \affiliation{\footnotesize Munich Centre for Quantum Science and Technology (MCQST), Schellingstra{\ss}e 4, 80799 M{\"u}nchen, Germany}
    \author{Sebastian Hohenemser}
    \affiliation{\footnotesize Technical University of Munich, TUM School of Natural Sciences, Department of Chemistry, Lichtenbergstra{\ss}e 4, 85748 Garching, Germany}
    \author{Florian Marquardt}
	\affiliation{\footnotesize Max Planck Institute for the Science of Light, Staudtstraße 2, 91058 Erlangen, Germany}
    \affiliation{\footnotesize Department of Physics, Friedrich-Alexander Universität Erlangen-Nürnberg, Staudtstraße 5, 91058 Erlangen, Germany}
	\author{Steffen J. Glaser}
	\email{glaser@tum.de}
	\affiliation{\footnotesize Technical University of Munich, TUM School of Natural Sciences, Department of Chemistry, Lichtenbergstra{\ss}e 4, 85748 Garching, Germany}
	\affiliation{\footnotesize Munich Centre for Quantum Science and Technology (MCQST), Schellingstra{\ss}e 4, 80799 M{\"u}nchen, Germany}

\begin{abstract}
Real-time adaptive control of quantum systems requires rapid generation of robust, high-fidelity pulses across a continuous range of operating conditions. Standard optimization algorithms such as gradient-ascent pulse engineering (GRAPE) solve each instance independently, discarding information between runs and requiring costly reinitialization when parameters change. We present an approach to robust optimal quantum control based on model-based reinforcement learning, in which a single neural network — embedding the Hamiltonian directly into the training pipeline — generates robust gates across an entire family of gate configurations, without pre-computed training data. Demonstrated on a single-spin (two-level) system, the trained networks produce pulses for arbitrary rotation angles over a range of pulse durations, detunings, and field inhomogeneities in milliseconds, at fidelities comparable to multi-seed GRAPE. The framework is inherently adaptable: any parameter entering the Hamiltonian can serve as a network input, extending the approach to different systems and control settings. Beyond speed, the network reveals structure in the control landscape: it discovers the same structured phase profiles that appear in GRAPE solutions — made identifiable through fidelity-invariant symmetry transformations — but more consistently than independent optimization. This consistency enables smooth interpolation across the entire trained parameter space.

\end{abstract}
\maketitle
\section{Introduction}
Precise control of quantum systems lies at the heart of modern quantum technologies. Manipulating qubit states with high fidelity requires carefully designed control pulses that steer the system dynamics towards a desired outcome, whether a specific state-to-state transfer or the realization of a target gate~\cite{ChristianeKoch_QOC2022,Ansel_QOC2024,Koch_OpenSystem}. However, in realistic experimental settings, the ideal control fields are distorted by various hardware limitations, such as finite amplitude constraints and inhomogeneities in qubit parameters such as frequency detuning~\cite{NMR_techniques_for_quantum_control_and_computation,Quantum_information_with_Rydberg_atoms,Spins_in_few_electron_quantum_dots}. These imperfections lead to systematic gate errors which ultimately degrade the performance of quantum processors~\cite{Preskill_NISQera}. Pulse optimization has emerged as a central tool in quantum control engineering to mitigate the errors and thereby increase the fidelity of quantum systems~\cite{Glaser_Review2015}. Consequently, many advanced experimental protocols—ranging from high-field NMR (nuclear magnetic resonance), parallel transmit MRI (magnetic resonance imaging) to pulse-level variational quantum eigensolvers (VQE) and continuous gate calibrations—require the on-the-fly optimization of tailored robust control pulses to account for specific hardware variations and continuous parameter spaces~\cite{VanDamme2021,glaser_real_fidelity_paper,GRAPE_paper,continous_quantum_gates_paper,Improving_the_performance_of_deep_quantum_optimization_algorithms_with_continuous_ate_sets,Gate_free_state_preparation_for_fast_variational_quantum_eigensolver_simulations}.\\
Techniques such as GRAPE (gradient ascent pulse engineering)~\cite{GRAPE_paper,glaser_real_fidelity_paper} have enabled the creation of control fields that achieve gate fidelities approaching the physical limits imposed by the experimental setup. In particular, these methods have proven essential in a wide range of hardware platforms, including superconducting qubits ~\cite{Optimal_control_of_coupled_Josephson_qubits,Implementing_a_universal_gate_set_on_a_logical_qubit_encoded_in_an_oscillator}, trapped ions~\cite{Jandura2022_Rydberg,Evered2023_ParallelRydberg}, NMR ~\cite{GRAPE_paper,DavidJoseph_NMR1p2GHz} and Nitrogen-vacancy (NV) centers ~\cite{High_fidelity_spin_entanglement_using_optimal_control}.\\
A known limitation of GRAPE is its inherently memoryless character. Each optimization run operates independently, without retaining information about previously optimized control fields or nearby solutions in parameter space. Consequently, even small perturbations of the system parameters can alter the control landscape sufficiently that GRAPE must be reinitialized and reoptimized from scratch, making it a poor candidate for on-the-fly optimization. Machine-learning–based approaches offer a way to mitigate this limitation by enabling a more global exploration of the control landscape. Rather than relying solely on local gradient information, such methods can learn adaptive representations or control policies that can escape unfavorable local minima and identify robust, high-fidelity solutions. Importantly, once trained, these models can also interpolate across different system parameters, allowing them to generalize solutions and provide high-quality initial guesses—or even direct predictions—without requiring a full optimization for every new parameter setting. Crucially, because inference requires only a forward computational pass, a trained network can synthesize complex pulses almost instantaneously—delivering direct on-the-fly predictions or high-quality initial guesses in milliseconds and effectively circumventing the prolonged execution times of iterative numerical solvers. \\
Supervised learning has been shown to work well~\cite{Vinding_paper,High_fidelity_control_of_spin_ensemble_dynamics_via_artificial_intelligence_from_quantum_computing_to_NMR_spectroscopy_and_imaging} but its performance is inherently bounded by the quality of the training data — it cannot surpass the solutions it was trained to reproduce. It relies on large training datasets generated by conventional optimization algorithms, whose computational cost makes data generation prohibitive for large systems. Model-Free Reinforcement Learning (RL) can discover novel strategies without prior data and has been shown to work well for problems where a physical model of the system is not available ~\cite{PhysRevX.8.031084,PhysRevX.8.031086,reuer2023realizing,bukov2026reinforcementlearningquantumtechnology,Experimental_Deep_Reinforcement_Learning_for_Error-Robust_Gate-Set_Design_on_a_Superconducting_Quantum_Computer,Model_Free_Quantum_Control_with_Reinforcement_Learning}. Because these agents are physics-agnostic, they must learn the governing dynamics entirely from reward signals, without access to analytical gradients of the physical objective. This causes severe sample inefficiency, wasteful exploration of physically irrelevant parameter spaces, and can lead to fragile solutions that lack experimental robustness ~\cite{Robust_quantum_control_using_reinforcement_learning_from_demonstration}.\\
When the system Hamiltonian is known, model-based reinforcement learning offers a natural alternative ~\cite{Schaefer_2020,FlorianMarquardt_cat_paper,PRXQuantum.4.030305}. Unlike conventional neural networks that rely on pre-generated training data, in a model-based RL approach, one can embed the physical objective directly into the training loop by constructing a fully differentiable pipeline from network output through the physics simulation to the loss function and use automatic differentiation (AD) for training~\cite{Schaefer_2020,PhysRevA.95.042318}.\\
In this work, we use model-based RL to present an approach to the optimal control of robust quantum gates and apply it to a single spin (two-level) system driven by an external field. We adopt the term model-based RL following~\cite{FlorianMarquardt_cat_paper,bukov2026reinforcementlearningquantumtechnology}. We demonstrate that the
resulting pulses achieve fidelities comparable to those
from multi-seed GRAPE optimization with neither method consistently
outperforming the other. Our investigation was motivated by the observation that GRAPE-optimized pulses occasionally exhibit structured phase patterns, despite stochastic initialization. Fidelity-invariant symmetry transformations make these patterns identifiable, suggesting the existence of distinct solution families in the control landscape. To explore this systematically, we
trained a neural network on the same control problem and
found that it consistently discovers these structured
solutions across the full parameter space. These
structures are preserved under subsequent GRAPE
refinement with little to no loss in fidelity.
\begin{figure}[h!]
    \centering
    \includegraphics[width=\columnwidth]{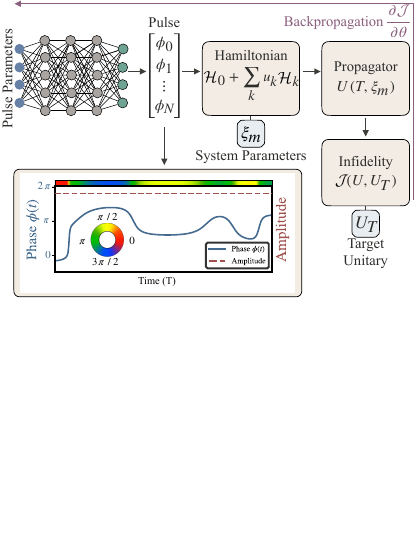}
    \vspace{-38ex}
    \caption{Schematic of model-based RL for quantum control. The network generates a control pulse, which is then applied to the system Hamiltonian to compute the resulting propagator. The infidelity with respect to the target unitary is evaluated and backpropagated through the entire computational graph using automatic differentiation (AD) to update the network parameters.}
    \label{fig:paper_sketch}
\end{figure}
\section{Model-based RL for Robust Quantum Control}
\label{sec: Model-based RL for Robust Quantum Control}
Deep learning and the GRAPE algorithm share a common
mathematical foundation: both methods try to minimize a cost
function by iteratively adjusting parameters in the
direction of steepest descent (ascent). These gradients
are obtained, in both cases, by systematic application of
the chain rule through the sequence of operations that
maps parameters to cost --- a procedure known as
backpropagation in the machine learning community and as
the co-state method in optimal control theory. These
analogies naturally motivate the application of optimal control
concepts within neural network
frameworks~\cite{Schaefer_2020,FlorianMarquardt_cat_paper,PRXQuantum.4.030305}.

In the framework presented here, we design robust control pulses
that maintain high fidelity despite experimental imperfections.

In general, the full system Hamiltonian (including the control Hamiltonian) can be written as
\begin{equation}
    \mathcal{H}(t,\bm{\xi}_m)= \mathcal H_0(\bm{\xi}_m)+ \sum_{k}u_k(t)\mathcal H_k(\bm{\xi}_m),
    \label{eq:general_control_ham}
\end{equation}
where $\mathcal{H}_0$ denotes the drift Hamiltonian, while $\mathcal{H}_k$ are the control Hamiltonians associated with the respective control fields $k$ and control amplitudes $u_k(t)$. $\bm{\xi}_m$ denotes the set of system parameters. Starting from an initial set of random weights and biases $\theta_i$ in the neural network, a corresponding control pulse is generated. To ensure robustness against experimental imperfections, this
pulse is applied to the Hamiltonian for $M$ different sets of
system parameters $\bm{\xi}_m$, sampled from a grid
spanning the expected range of experimental variations. For each
parameter set, a quantum simulation constructs the total propagator
%
\begin{equation}
    U(T,\bm{\xi}_m) = U_N(\bm{\xi}_m)\,U_{N-1}(\bm{\xi}_m)\cdots U_2(\bm{\xi}_m)\,U_1(\bm{\xi}_m)
\end{equation}
with $U_j(\bm{\xi}_m) = \exp\left(-i\mathcal{H}_j(\bm{\xi}_m) \Delta t\right)$ and the Hamiltonian is
piecewise constant within each time step. These propagators are then evaluated against the desired
target operator $U_T$ to obtain the ensemble-averaged infidelity
\begin{equation}
    \mathcal{J} = 1 - \frac{1}{dM}\sum_{m=1}^{M}
    \mathrm{Re}\left\{\mathrm{Tr}(U_T^\dagger U(T,\bm{\xi}_m))\right\},
    \label{eq:infidelity}
\end{equation}
where $d=2$ is the dimension of the Hilbert space. By averaging the fidelity over this ensemble, the optimization
is driven toward control pulses that perform well not only under ideal conditions (no experimental errors) but across the full range of
parameter variations. The global-phase-sensitive form of the fidelity metric is particularly important for large robustness ranges. By retaining global-phase information it specifies a single well-defined target propagator at each parameter value, so that gradients averaged over the robustness range reinforce one another rather than converging toward conflicting global-phase-equivalent targets, substantially increasing the probability of reaching high-fidelity solutions~\cite{GRAPE_paper,glaser_real_fidelity_paper,JANICH2011126}. We note that while the optimization is performed with a global-phase-sensitive cost function, experimental observables are insensitive to global phase. A pulse optimized for $R_y(\beta)$ and a pulse independently optimized for $R^{ps}_{y}(2\pi - \beta)$ — where $ps$ denotes a phase shift of all pulse phases by $\pi$ — implement the same quantum gate in experiment. Since these are independent optimization tasks, they yield different pulses and fidelities. Optimizing over rotation angles $\beta \in [0, 2\pi]$ therefore provides access to all y-axis rotation gates, with complementary angles related by a global phase shift. 

In order for the network to generalize, we train on
batches --- sets of $B$ randomly sampled gate
specifications from the target parameter space $\mathcal{G}$---
rather than optimizing one pulse at a time. During each
backpropagation step the network processes the average
infidelity $\bar{\mathcal{J}}$ computed over the batch. The network learns the structure of optimal pulses by
calculating the gradient through the entire chain from the batch-averaged
infidelity $\bar{\mathcal{J}}$, through the propagators, through
the controls $u_{k,j}$, to each weight and bias $\theta_i$.

\subsection{Gradient structure and connection to GRAPE}
In GRAPE, the gradient of the infidelity with respect to the
control field $u_{k,j}$ at time step $j$ is obtained through a
forward--backward decomposition of the propagator chain. Denoting
the propagator for a single time step as
$P_j^{(m)} = \exp(-i\,\mathcal{H}_j(\bm{\xi}_m)\,
\Delta t)$, the gradient reads
\begin{equation}
    \frac{\partial \mathcal{J}}{\partial u_{k,j}}
    = -\frac{1}{M}\sum_{m=1}^{M}\mathrm{Re}\!\left\{
      \mathrm{Tr}\!\left(
        B_{j+1}^{(m)\dagger}\,
        \frac{\partial P_j^{(m)}}{\partial u_{k,j}}\,
        F_{j-1}^{(m)}
      \right)\right\},
    \label{eq:grape_gradient}
\end{equation}
where $F_{j-1}^{(m)} = P_{j-1}^{(m)} \cdots P_1^{(m)}$ is the
forward-propagated state up to step $j{-}1$ and
$B_{j+1}^{(m)} = P_j^{(m)\dagger} \cdots P_N^{(m)\dagger}\,U_T$
is the target operator back-propagated to step $j{+}1$. The
gradient at each time step is thus determined by the overlap
between the forward state (encoding the evolution up to that
point) and the backward costate (encoding what the target
requires from that point onward), connected through the local
propagator derivative $\partial P_j^{(m)} / \partial u_{k,j}$.
All $N$ gradients are obtained in a single forward--backward pass
over the propagator chain---the same algorithmic structure that
automatic differentiation produces when applying the chain rule
to a sequential product of matrices. 

In the neural network framework, the controls are no longer free
parameters but functions of the shared network weights:
$u_{k,j}^{(b)} = f(\theta, \mathbf{x}_b)_{k,j}$, where
$\mathbf{x}_b$ specifies the $b$-th gate in a batch of $B$ gate
specifications. The gradient with respect to the network weights
follows by the chain rule:
\begin{equation}
    \frac{\mathrm{d}\bar{\mathcal{J}}}{\mathrm{d}\theta_i}
    = \frac{1}{B}\sum_{b=1}^{B}\sum_{k,j}
      \frac{\partial \mathcal{J}_b}
        {\partial u_{k,j}^{(b)}}
      \frac{\partial u_{k,j}^{(b)}}
        {\partial \theta_i}.
    \label{eq:backprop}
\end{equation}
The first factor is identical to the GRAPE gradient of
Eq.~\eqref{eq:grape_gradient}, averaged over the robustness
ensemble. The second factor is the Jacobian of the network
mapping from weights to control fields, which is absent in
GRAPE. In practice, automatic differentiation evaluates the
entire chain---from infidelity through the propagator sequence
and continuing through the network layers to the weights---without
requiring a manual derivation of the backward pass.

Despite this structural similarity, the scope of each update is
fundamentally different. In GRAPE, each gradient step optimizes a
single pulse in isolation: updating $u_{k,j}$ affects only one
control field at one time step of one pulse. In this framework, each update to the weights and biases $\theta_i$ simultaneously improves pulses across the entire batch,
because all gate specifications share the same network weights. This is what
enables the network to generalize across the parameter space---a
property that GRAPE, by construction, cannot possess.

\section{NN Architecture for Robust Optimal Control Pulses}
\begin{figure}
    \centering
    \includegraphics[width=\columnwidth]{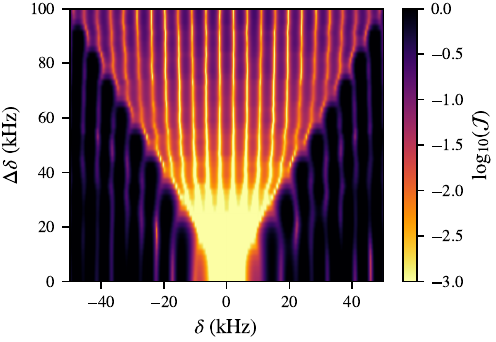}
    \caption{Robustness profile of pulses with $T_\text{P} = 450\,\mu\mathrm{s}$, $\beta=3\pi/2$ and $\Delta s = 0$ optimized for different offset ranges $\Delta\delta$.}
    \label{fig: robustness_illustration}
\end{figure}
In this section, we explain the architecture used to generate robust quantum gates with neural networks and its application on a single-qubit control problem. The central idea is to amortize the optimization: rather than solving a separate optimal-control problem for every gate configuration, as conventional methods do, we train a single network that maps an entire continuous family of configurations to their optimized pulses at once. Once trained, the network produces a high-fidelity pulse for any configuration in this space, including those never seen during training, in milliseconds and without re-optimization. 

We use a simple fully connected neural network.
Physical parameters that do not change the dimensionality of the network
output, such as the rotation angle or the (constant) pulse amplitude,
enter naturally as network inputs: optimizing over a range of such a
parameter requires only that it be sampled and supplied as an additional
input. The pulse duration is an exception, since it sets the number of
piecewise-constant time slices. One option is a fixed-dimensional
parametrization --- a spline or Fourier basis --- whose coefficient count
is independent of duration, or alternatively a fixed number of slices with
a duration-dependent step $dt$. We instead keep $dt$ fixed and optimize the
phase of every slice independently, which spans the full piecewise-constant
solution space rather than a restricted basis and thus admits the
highest-fidelity solutions. To accommodate variable duration within a
fixed-size output layer, we employ a masking scheme: for a pulse of $L$
time slices, only $L$ of the $N_{\max}$ output neurons carry active phase
values, while the remaining $N_{\max}-L$ entries are set to zero so that
their gradients do not contribute during backpropagation.

Piecewise-constant optimization can produce high-frequency phase
oscillations in the network output that are difficult to implement on
experimental hardware. We therefore apply a smoothing filter, subject to
three requirements: it must be fully differentiable to allow gradient
propagation during training, introduce no phase lag so as to preserve the
temporal structure of the pulse, and add minimal computational overhead per
training step. The bidirectional exponential moving average (EMA) satisfies
all three with a single hyperparameter. For a raw phase sequence
$\{\phi_j\}$, the smoothed sequence $\{\tilde{\phi}_j\}$ is obtained by a
forward pass
\begin{equation}
    f_t = \epsilon\,\phi_t + (1 - \epsilon)\,f_{t-1},
\end{equation}
followed by a backward pass over the reversed sequence,
\begin{equation}
    \tilde{\phi}_t = \epsilon\,f_t + (1 - \epsilon)\,\tilde{\phi}_{t+1},
\end{equation}
where $\epsilon \in (0,1]$ is the smoothing factor and the bidirectional
application cancels the phase lag of a single pass.

Unlike supervised learning, whose performance is ultimately bounded by the
size of the labeled dataset, we sample gate configurations randomly from the gate configuration space
$\mathcal{G}$ in every batch, providing an effectively infinite training
set. The network therefore can learn the underlying structure of optimal pulse
design rather than interpolating a fixed set of precomputed solutions.

Finally, because the configurations in $\mathcal{G}$ can span a wide range of
achievable fidelities, loss scheduling improves training across this
spread. Starting from the infidelity $\mathcal{J}$ of
Eq.~\eqref{eq:infidelity}, one may reweight the loss --- for example using
$\log\mathcal{J}$, or a modified loss $\mathcal{J}^{\alpha}$ with
$\alpha < 1$ --- to increase the relative weight of high-fidelity
configurations in the late stages of training, encouraging the network to
further polish its best solutions.

We apply this framework to the optimal control of robust single-qubit rotation gates. The network outputs a time-dependent phase profile $\phi(t)$ at constant (nominal) amplitude $\nu$. This reflects typical RF hardware constraints and, in the amplitude-limited regime relevant here, is not expected to limit performance: when a free amplitude degree of freedom is allowed, optimal broadband solutions saturate the amplitude bound imposed during optimization, recovering the constant-amplitude regime~\cite{glaser_real_fidelity_paper}. The control Hamiltonian includes two
dominant experimental inhomogeneities: qubit
frequency detuning and amplitude miscalibration of the control
field. The system dynamics are governed by
\begin{equation}
    \mathcal H(t)
    = 2\pi\Bigl[
        \delta\, I_z
        + s\nu\bigl(
            I_x \cos(\phi(t)) + I_y \sin(\phi(t))
        \bigr)
    \Bigr],
    \label{eq:1spin_control_hamiltonian}
\end{equation}
where $I_{x,y,z} = \sigma_{x,y,z}/2$ are the spin-1/2 operators,
$\delta$ denotes a deviation of the qubit transition frequency from
the nominal drive frequency, and $s\cdot\nu$
represents the actual Rabi rate including a systematic amplitude
error with $s$ being the amplitude scaling factor. This Hamiltonian describes a driven two-level system in the
rotating frame within the rotating-wave approximation (RWA), a model applicable to single-qubit control across many
experimental platforms whenever the two-level approximation
is valid~\cite{Glaser_Review2015,ChristianeKoch_QOC2022,Ansel_QOC2024}. The control amplitudes are constructed from the network
output as $u_x(t) = s\cdot\nu\cos\phi(t)$ and
$u_y(t) = s\cdot\nu\sin\phi(t)$, recovering the
structure of Eq.~\eqref{eq:general_control_ham}.

We use center-aligned masking, placing the $L$ active outputs symmetrically about the midpoint of the output layer. The alignment has a small to negligible effect on fidelity compared to left- or right-aligned masking; we adopt the symmetric choice as it is structurally compatible with the phase profiles we found in GRAPE optimized pulses in section~\ref{sec:structure}. We found $\mathcal{J}^\alpha$ ($\alpha<1$) to work better for loss scheduling than $\log\mathcal{J}$.

\begin{figure*}[t]
  \includegraphics[width=\textwidth]{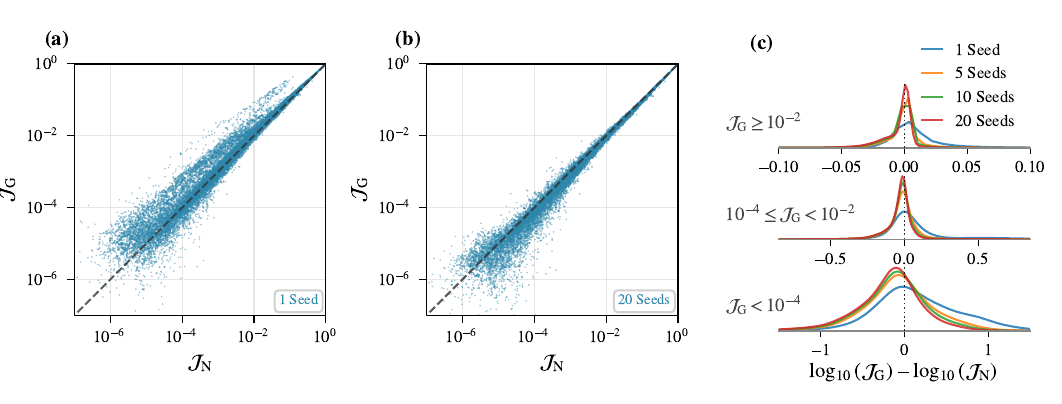}
    \caption{Comparison of neural network and GRAPE infidelities
  across 100{,}000 gate configurations $\{G_i\}$.
  (a,\,b)~Infidelity of GRAPE ($\mathcal{J}_\mathrm{G}$)
  versus the neural network ($\mathcal{J}_\mathrm{N}$) for
  1 and 20 GRAPE seeds, respectively; each point represents
  one gate configuration within $\{G_i\}$ and the dashed line indicates equal
  performance. The network outperforms single-seed GRAPE in a substantial fraction of cases; with 20 seeds the balance shifts in favor of GRAPE, though most configurations remain near the diagonal.
  (c)~Distribution of
  $\log_{10}\mathcal{J}_\mathrm{G} -
  \log_{10}\mathcal{J}_\mathrm{N}$ grouped by infidelity
  regime (rows) and GRAPE seed count (colors). Negative values
  indicate GRAPE outperforms the network. The distributions
  broaden and shift toward negative values with increasing seed
  count and decreasing infidelity.}
    \label{fig:nn_analysis}
\end{figure*}

The network takes four inputs, defined as a gate configuration $G_i = (\beta,\, T_\text{P},\, \Delta\delta,\, \Delta s) \in \mathcal{G}$: rotation angle, pulse duration, target detuning range, and target amplitude inhomogeneity range. Here, $\Delta\delta$ and $\Delta s$ denote the widths of symmetric robustness intervals: 
the pulse is required to maintain high fidelity over 
$\delta \in [-\Delta\delta/2, +\Delta\delta/2]$ and $s \in [1-\Delta s/2, 1+\Delta s/2]$, respectively. All target rotations are rotations around the y-axis. Rotation angles and pulse durations are sampled from discrete sets
during training. Two networks with identical architecture are trained: one covering $\beta \in [0,\pi]$ and one covering $\beta \in (\pi,2\pi]$, each with rotation angles sampled in increments of $\pi/8$. A single network over $[0,2\pi]$ achieves comparable fidelities, but the split yields smoother interpolation, consistent with the abrupt structural change of optimal pulses near $\beta=\pi$ (Sec.~\ref{sec:structure}). To stabilize the second network immediately above the domain boundary, it additionally includes an anchor point one degree above $\pi$, i.e. $\beta = \pi + \pi/180$ ($181^\circ$). Pulse durations are sampled from $5$ to
$450\,\mu\mathrm{s}$ in steps of $5\,\mu\mathrm{s}$ for both networks. Detuning and amplitude
inhomogeneity ranges are sampled continuously up to
$100\,\mathrm{kHz}$ and $40\%$, a regime where numerical
simulations faithfully reproduce experimental
performance~\cite{glaser_real_fidelity_paper,DavidJoseph_NMR1p2GHz}. We sample $\beta$ and $T_\text{P}$ from discrete grids rather than continuously. Under matched architecture and training budget, discrete sampling reached systematically lower infidelities than continuous sampling of both parameters. The network nonetheless generalizes to the continuum, interpolating smoothly to unsampled $(\beta, T_\text{P})$. Discrete anchors thus stabilize training while interpolation recovers continuous coverage. While we demonstrate the framework with four input parameters, the architecture imposes no restriction on the dimensionality of
$\mathcal{G}$: it can be extended to include additional parameters — such as a variable RF amplitude or spline-based pulse parameterizations — without changing the training procedure. Once trained, the network can accept any input values within the training parameter space. Figure~\ref{fig: robustness_illustration} illustrates how the
robustness specification directly shapes the optimized pulse. Each
row shows $\mathcal{J}(\delta)$ of a neural network-generated pulse with $T_\text{P} = 450\,\mu
\text{s}$ and $\Delta s = 0$, optimized for a different target
offset range $\Delta\delta=0\,\text{kHz}$ to $\Delta\delta=100\,\text{kHz}$. These optimized pulses achieve high fidelities within their target window
but deteriorate rapidly outside it. Pulses optimized for broad offset ranges achieve high fidelities across the full window, while those targeting narrower ranges can concentrate their performance budget and reach significantly higher fidelities — all at the cost of rapid deterioration outside the target window.
The figure also reveals a fine vertical stripe pattern: narrow high-fidelity windows at regularly spaced offset values. These stripes are a genuine physical feature, not a sampling artifact: the infidelity is evaluated on a 2001-point grid, far finer than the stripe spacing, and the optimization itself uses 101 detuning points — which is also denser than the stripe spacing. Repeating the optimization with 201 detuning points reproduces the same pattern. We found that the spacing of high-fidelity vertical lines is roughly $2/T_P$ independent of rotation angle.
This illustrates the need for individually tailored pulses for
each robustness specification — precisely the task that the
neural network is trained to perform across the full gate
configuration space $\mathcal{G}$. This enables on-the-fly pulse generation and allows different robustness margins to be applied to individual qubits within the same device.

\begin{table}
  \caption{Neural network architecture and training hyperparameters.
  Two networks with this architecture are trained, one for
$\beta \in [0,\pi]$ and one for $\beta \in (\pi, 2\pi]$; each maps
four input parameters to a discretized phase profile of
$N_{\max} = 900$ time slices. Training proceeds in two
  stages: an initial training phase followed by a fine-tuning phase
  at reduced learning rate.}
\label{tab:architecture}
\begin{ruledtabular}
\begin{tabular}{lcc}
    Layer          & Neurons          & Activation \\
\hline
    Input          & 4                & ---        \\
    Hidden 1--5    & 2048             & ReLU       \\
    Output         & 900              & Linear     \\
\hline
\multicolumn{3}{l}{Total trainable parameters:
$\approx 18 \times 10^6$} \\[4pt]
\hline
                   & Initial training & Fine-tuning \\
\hline
    Learning rate
                   & $2 \times 10^{-4}$
                   & $5 \times 10^{-6}$  \\
    Batch size     & \multicolumn{2}{c}{128}                 \\
    Optimizer      & \multicolumn{2}{c}{Adan}                \\
    EMA $\epsilon$   & \multicolumn{2}{c}{0.2 (32 unrolls)} \\
    Loss function  & $\mathcal{J}$
                   & $\mathcal{J}^{\alpha}$, $\alpha = 0.2$  \\
\end{tabular}
\end{ruledtabular}
\end{table}

The network architecture and training hyperparameters, summarized in Table~\ref{tab:architecture}, were selected through a combination of systematic sweeps and manual exploration (see Appendix~\ref{app:model_size} for details). Training proceeds in two stages: an
initial phase at a learning rate of $2\times 10^{-4}$, followed by
fine-tuning at $5 \times 10^{-6}$ (see Table~\ref{tab:architecture} for details). During the fine-tuning stage, the loss is replaced by
$\mathcal{J}^\alpha$ with $\alpha = 0.2$, which
increases the gradient weight on configurations that have
already reached high fidelity, encouraging the network to
polish its best solutions further. We use the Adan optimizer~\cite{xie2024adanadaptivenesterovmomentum}, which we found to converge faster than Adam~\cite{kingma2017adammethodstochasticoptimization}, the default optimizer in most machine learning applications. The network uses Rectified Linear Unit (ReLU) activations. The entire training pipeline, including the quantum simulation and backpropagation, is implemented in JAX~\cite{jax2018github}, which provides AD through the physics layer. Both the neural network and GRAPE pulses were optimized on a robustness grid of 101 detuning points and 5 amplitude inhomogeneity points, over which the propagators are computed and the infidelity is averaged.

During evaluation, GRAPE and the quantum simulation following the neural network forward pass evaluate the ensemble-averaged infidelity over a fine grid of
$2001 \times 21$  points in detuning and amplitude inhomogeneity to ensure a fair comparison. Figures~\ref{fig:nn_analysis}(a,\,b) compare the infidelities achieved by the neural network ($\mathcal{J}_\mathrm{N}$) and
GRAPE ($\mathcal{J}_\mathrm{G}$) across 100{,}000 randomly
sampled gate configurations with $\beta$ and $T_\text{P}$ drawn from the discrete training grid and $\Delta\delta,\Delta s$ sampled continuously within the training bounds, for 1 and 20 random GRAPE seeds. Each point
represents an individual gate configuration $G_i$; the dashed
line indicates equal performance. Compared to a single GRAPE seed, the neural network produces the better pulse in a substantial fraction of cases. As the number of seeds increases to 20, this fraction decreases, as multi-seed GRAPE explores a larger portion of the control landscape. 

To resolve the fine structure of this comparison,
Fig.~\ref{fig:nn_analysis}(c) shows the distribution of
$\log_{10}\mathcal{J}_\mathrm{G} -
\log_{10}\mathcal{J}_\mathrm{N}$ for different infidelity regimes
(rows) and GRAPE seed counts (colors). Negative values on the x-axis indicate GRAPE outperforms the neural network. At infidelities above $10^{-2}$,
the distributions are narrow and centered near zero, confirming
that both methods find similar solutions when the optimization
landscape is simple. As the infidelity decreases below $10^{-4}$,
the distributions broaden and shift toward negative values with
increasing seed count, indicating that GRAPE benefits from
additional random restarts in a regime where the optimization
landscape contains more local minima. Nevertheless, even at
$\mathcal{J}_\mathrm{G} < 10^{-4}$, the difference between both methods remains
within one order of magnitude. Remarkably, the network not only produces pulses in milliseconds but achieves fidelities competitive with GRAPE across the full target parameter space.

\begin{figure}
    \includegraphics[width=\columnwidth]{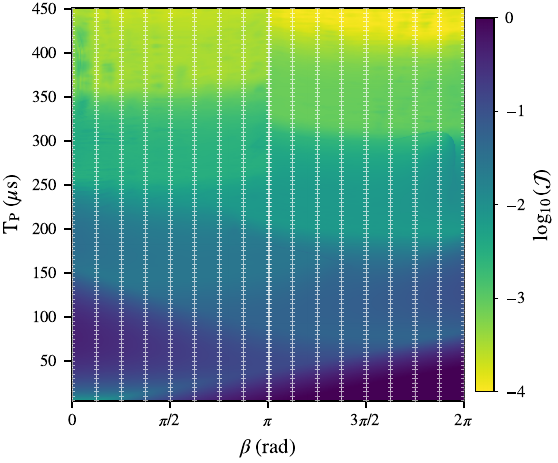}
    \caption{Infidelity landscape of pulses generated by the neural network, evaluated across a dense grid of pulse durations $T_P$ and rotation angles $\beta$ for $\Delta\delta = 20\,\mathrm{kHz}$ and $\Delta s=0$. White crosses indicate training points included in the training parameter space.}
    \label{fig:interpolation}
\end{figure}

Figure~\ref{fig:interpolation} demonstrates the networks' ability to
generalize beyond the discrete training grid. The trained models are
evaluated at $\Delta\delta=20\,\mathrm{kHz}$ and $\Delta s=0$ on a dense grid of rotation angles at $1^\circ$ step size
and pulse durations in steps of $1\,\mu\mathrm{s}$, the vast majority of which correspond to
$(\beta,\, T_\mathrm{P})$ combinations that cannot be seen during training (white
crosses mark points in the training parameter space). The infidelity landscape varies smoothly across both dimensions, confirming that the networks are able to generalize continuously beyond the discrete training grid.

Each network was trained on a single NVIDIA RTX 5090 consumer GPU, completing in 5 days and 1 hour per run, compared to 1 day and 4 hours for multi-seed GRAPE with 20 seeds parallelized across 250 threads on a dual-socket AMD EPYC 9554 server node. These wall-clock times are not directly comparable (see Appendix~\ref{app:computation}); however, the two optimizations differ fundamentally in scope: unlike GRAPE, which must be reoptimized independently for each new parameter setting, the trained networks together cover the full parameter space — any combination of rotation angle, pulse duration, detuning range, and amplitude inhomogeneity within the training bounds — generating any individual high-fidelity pulse in milliseconds without retraining. Generating an equivalent coverage with GRAPE is computationally intractable. The neural network generates the full dataset of 100{,}000 pulses, shown in Fig.~\ref{fig:nn_analysis} in seconds. 

\begin{figure}[h]
    \includegraphics[width=\columnwidth]{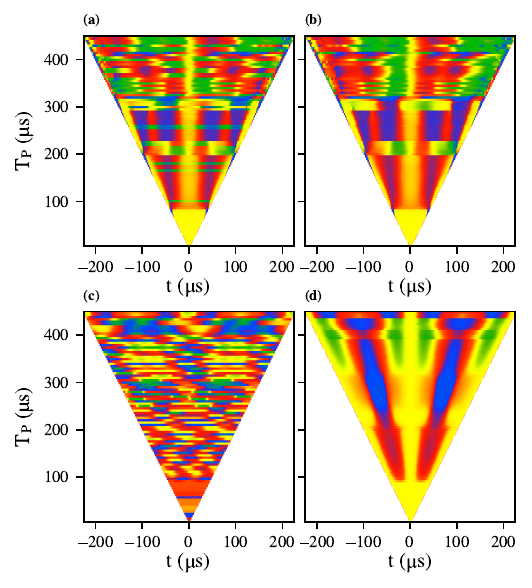}
    \caption{Phase profiles of GRAPE-optimized pulses before (left column) and
    after (right column) phase curation, for rotation angles $\beta = 3\pi/2$
    (top row) and $\beta = 2\pi$ (bottom row). The
    curation procedure reveals continuous phase structures while leaving the gate fidelity
    unchanged.}
    \label{fig: 2x2_cleaned_phase_profiles}
\end{figure}
\section{Structural Analysis of Optimized Pulses}
\label{sec:structure}
\begin{figure*} 
    \includegraphics[width=\linewidth]{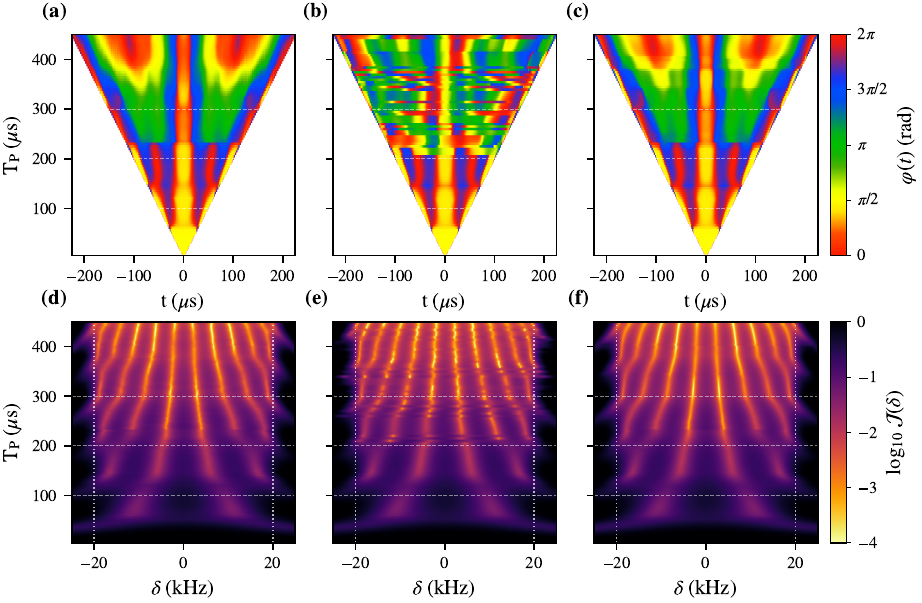}
    \caption{Optimized phase profiles and robustness fingerprints
    as a function of pulse duration $T_\text{P}$ for a $3\pi/2$
    rotation with $\Delta\delta = 40\,\mathrm{kHz}$ and
    $\Delta s = 0$. Top row: phase profiles $\varphi(t)$
    for (a)~the neural network, (b)~GRAPE (20 seeds, best
    retained), and (c)~GRAPE initialized from the neural network
    solution. The network produces smooth, symmetric
    phase patterns that persist across all durations. GRAPE mostly finds the same structures for short pulses.
    GRAPE initialized from the network refines within the same
    structured family, confirming these solutions correspond to
    genuine minima in the control landscape. Bottom row:
    robustness fingerprints showing
    $\log_{10}(\mathcal{J}(\delta))$, for (d)~NN,
    (e)~GRAPE, and (f)~NN$\to$GRAPE. Despite the differences in
    pulse structures, all three methods produce nearly identical
    robustness patterns.}
    \label{fig:structure_fingerprint}
\end{figure*}

Systematic analysis of GRAPE pulses reveals that the seemingly random phase profile functions $\Phi(t,\, T_\text{P},\, \Delta\delta,\, \Delta s,\, \beta)$ carry more structure than commonly assumed~\cite{KUPROV2013107}: These structures can be identified through an alignment process (\textit{curation}) which uses fidelity-invariant symmetry transformations (see Appendix~\ref{app:symmetry})~\cite{Braun_2014}. However, small changes in gate parameters can shift the optimum solution found by GRAPE to a different phase structure family. Figure~\ref{fig: 2x2_cleaned_phase_profiles} demonstrates the
effect of these curation procedures. Panels~(a) and~(b) show the uncurated and curated phase profiles for $\beta = 3\pi/2$, panels~(c) and~(d) for $\beta = 2\pi$, each as a function of pulse duration $T_\text{P}$ with all remaining gate parameters held fixed. In both cases, the
curation reveals continuous structural patterns across pulse
durations that are obscured in the raw GRAPE output. The effect is particularly pronounced for
$2\pi$ rotations.
This raises the question whether these continuous structures can be revealed systematically while remaining competitive in fidelity. The trained neural networks provide a natural
tool to address this question. Because the networks map all gate configurations through shared
weights, they are biased toward finding solutions that vary smoothly
across the parameter space. If the control landscape contains structured solution families that are competitive in fidelity, the networks will preferentially converge to them, since nearby gate configurations naturally produce similar pulses — a continuity that GRAPE, optimizing each pulse in isolation from random initialization, has no mechanism to exploit.

\begin{figure}
    \includegraphics{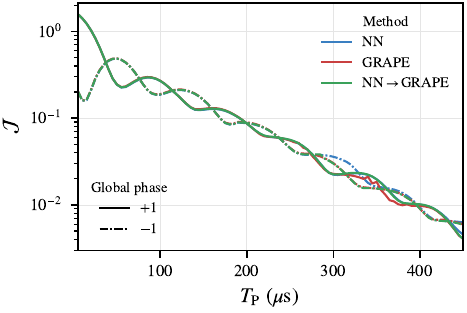}
    \caption{Infidelity as a function of pulse duration $T_\text{P}$ for a
  $\pi/2$ and $3\pi/2$ rotation. Note that both rotations (global phases $+1$ and $-1$) implement the same quantum gate in experiment. All three methods converge to essentially the same infidelity for
  $T < 300\,\mu\mathrm{s}$. At longer pulse durations, GRAPE achieves the lowest infidelity, followed by GRAPE initialized from the NN and the NN alone, though the differences remain marginal.}
    \label{fig:symm_loss}
\end{figure}

We illustrate the structural analysis for the case of a $3\pi/2$ target rotation
with $\Delta\delta = 40\,\mathrm{kHz}$ and
$\Delta s = 0$ as a representative example. This rotation
angle exhibits fewer symmetries than special cases such as $\{0$, $\pi$, $2\pi\}$, while still producing clearly structured phase
profiles. Figure~\ref{fig:structure_fingerprint}(a--c) compares the pulse
phase profiles generated by three approaches across the full
range of gate durations $T_\text{P}$ for a $3\pi/2$ rotation: the neural
network, GRAPE (20 random seeds, best result
retained), and GRAPE initialized from the neural network
solution. For short pulse durations, all three methods produce
qualitatively and quantitatively the same phase profiles, confirming that they
converge to compatible regions of the control landscape. As the
pulse duration increases, the GRAPE solutions
[Fig.~\ref{fig:structure_fingerprint}(b)] begin to alternate between different solution families that are absent at shorter durations. This is not because individual
GRAPE pulses lack structure, but rather because at each duration
the pulse with the lowest infidelity is selected, which may belong to a different structural family. This selection by fidelity alone, without regard for
structural continuity, produces more unstructured phase profiles
visible across the parameter space. As Figure~\ref{fig:symm_loss} shows, the fidelity difference between structured and unstructured solutions is often negligible. This makes the structured pulses a practical alternative, with the additional advantage of being amenable to smooth interpolation across the parameter space. The neural network
(Fig.~\ref{fig:structure_fingerprint}(a)) produces consistently structured phase profiles across the full range of pulse durations, making the underlying patterns clearly visible. The most informative comparison is
Fig.~\ref{fig:structure_fingerprint}(c): when GRAPE is initialized
with the neural network pulse rather than from random seeds, the
resulting solution largely retains the structured phase profile of the
network output. Rather than departing from this structure and
converging to a solution resembling panel~(b), GRAPE refines
within the same region of the control landscape. Although the short pulses of panel~(a) appear visually distinct from their counterparts in panels~(b) and~(c), the curation procedure of Appendix~\ref{app:symmetry} maps one representation into the other without altering the fidelity.

To compare the physical performance of the different solution
families, Fig.~\ref{fig:structure_fingerprint}(d--f) shows
\textit{robustness fingerprints} for all three methods. For each pulse
duration and method, the offset-dependent infidelity is plotted
as $\log_{10}(\mathcal{J}(\delta))$.
Despite the visible differences in pulse structures, all three
methods produce nearly identical fingerprints, indicating that
the robustness features are largely independent of the specific
pulse structures.

Independent analysis of individual GRAPE pulses confirms that
structured solutions are not unique to the neural network:
depending on the rotation angle, GRAPE pulses frequently exhibit
symmetry or antisymmetry of their phase profiles
about the pulse midpoint~\cite{glaser_real_fidelity_paper}. This
mirror-like structure is also clearly visible in the neural
network solutions of Fig.~\ref{fig:structure_fingerprint}(a),
where the phase patterns are symmetric about the
temporal center of each pulse.

Figure~\ref{fig:symm_loss} compares the infidelity of all three
methods as a function of pulse duration, showing results for both
$\pi/2$ and $3\pi/2$ rotations. These pulses implement
the same quantum gate up to a global phase and a phase shift of $\pi$ of the $\beta=3\pi/2$ pulse and are therefore
physically equivalent in experiment. Because the infidelity measure
used here is sensitive to global phase, however, the two cases are
treated as distinct optimization targets and optimized individually.
This leads to an overall improvement in the gate fidelity that is
relevant for experimental practice.
For $T < 300\,\mu\mathrm{s}$, all three methods
converge to essentially the same infidelity, consistent with the
observation that at short durations the control landscape is
sufficiently constrained that all methods find the same solution
family. At longer durations, multi-seed GRAPE achieves the lowest infidelity, followed by GRAPE initialized using the NN and the network alone, though all three remain within close range of each other.

This ordering is expected: GRAPE optimizes each pulse individually
to full convergence and selects the best result from multiple
seeds, whereas the network must compromise across the entire
parameter space through shared weights. When GRAPE is initialized
from the network solution, it refines within the same structured
family rather than converging to the unrelated solutions of
panel~(b), while still achieving excellent (almost optimal) infidelities --- requiring only a single GRAPE run from
the network initialization rather than 20 independent seeds. Even
without this refinement step, the network alone produces pulses of
comparable performance across the full range of pulse durations in
a single forward pass requiring only milliseconds of computation.

\begin{figure*}[t]
    \includegraphics[width=\textwidth]{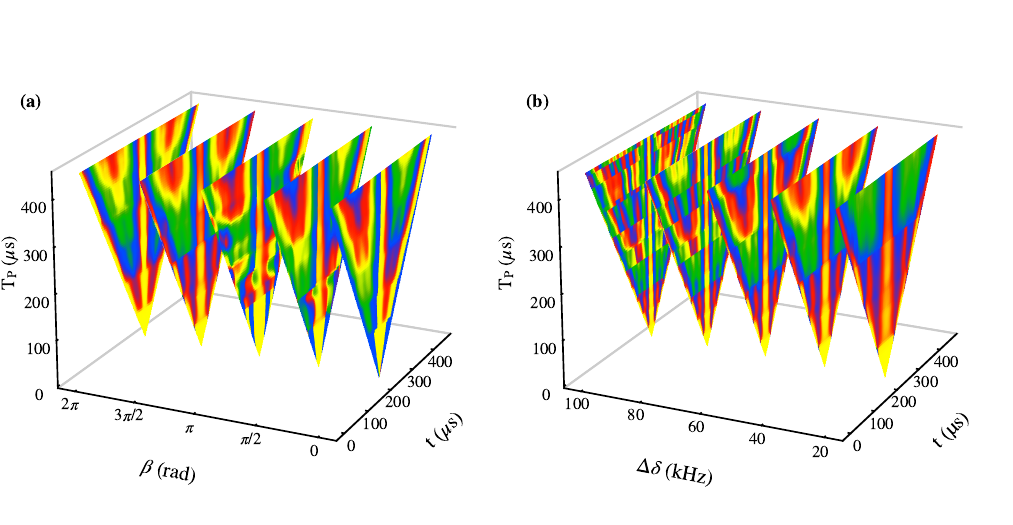}
    \caption{Neural-network-predicted pulses as a function of
(a) rotation angle $\beta$ and pulse duration $T_\mathrm{P}$, (b) offset range $\Delta\delta$ and pulse duration $T_\mathrm{P}$. Each slice shows the phase $\phi(t, T_\mathrm{P})$
of the optimized pulse sequence for different $\beta$ or $\Delta\delta$ respectively with $\Delta s=0$.}
    \label{fig:stacked_plots_offset}
\end{figure*}

Figure~\ref{fig:stacked_plots_offset} illustrates the capability of the network to generate
a diverse set of pulse
profiles across
dimensions of the gate configuration space $\mathcal{G}$ in milliseconds.
Panel~(a) shows the phase profiles $\varphi(t, T_\text{P})$ for five
rotation angles $\beta$ from $0$ to $2\pi$, evaluated using both
networks across their respective domains ($\beta \leq \pi$ and
$\beta > \pi$) at fixed
offset range $\Delta\delta =  40\,\mathrm{kHz}$ and
$\Delta s = 0$. The pulse structure varies substantially with
rotation angle: the $0$, $3\pi/2$ and $2\pi$ slices always exhibit symmetric or antisymmetric
patterns, while for intermediate angles this is not the case. Panel~(b) shows
how the phase profiles evolve as the target offset range
$\Delta\delta$ increases from $0$ to $100\,\mathrm{kHz}$ at
fixed $\beta = 3\pi/2$ and $\Delta s = 0$. For narrow offset
ranges, the pulses are relatively simple; as the robustness
requirement grows, the phase profiles develop increasingly rich
structure to compensate for the larger range of detuning errors.
In both panels, each slice represents a full triangle plot of
$\varphi(t)$ versus $T_\text{P}$, generated in a single forward
pass of the network. The smooth variation between neighboring
slices reflects the network's ability to interpolate continuously
across the parameter space.

\section{Conclusion}

We have presented a model-based RL framework for the
optimal control of robust quantum gates and applied it on a single-qubit problem across a multi-dimensional parameter space. By embedding the system
Hamiltonian directly into the training pipeline, the network learns
to produce high-fidelity control pulses for gate
configurations — spanning pulse duration, rotation angle,
frequency detuning, and amplitude inhomogeneity — in a single
forward pass requiring only milliseconds per pulse. This enables
the on-the-fly generation of individually adapted pulses. We demonstrated this on a single-qubit
Hamiltonian with NMR-realistic parameters. Benchmarked against multi-seed GRAPE on the training parameter space, the network achieves
comparable fidelities — with neither method consistently outperforming
the other — while additionally supporting high-fidelity continuous interpolation across the parameter space without reoptimization. The network
routinely achieves infidelities below $10^{-3}$ — well within
the requirements of typical NMR and MRI applications~\cite{glaser_real_fidelity_paper}.
Covering this space densely with GRAPE would require an independent optimization for each parameter combination — computationally intractable at the scale evaluated here. For
platforms demanding higher precision, such as gate calibration in
quantum computing, the same framework can be straightforwardly
adapted by changing the Hamiltonian, narrowing the robustness parameter space or increasing
the training effort, trading generality for fidelity in the
regime of interest.

Beyond computational efficiency, our results provide novel insight into
the structure of the quantum optimal control landscape. These structured solutions
were first identified in GRAPE through a curation procedure using fidelity-invariant symmetry transformations,
and independently confirmed by the neural network. When GRAPE is initialized from neural network solutions, it refines within the same structured family, and the resulting robustness fingerprints are nearly identical across all three methods — confirming that qualitatively different phase profiles can achieve very similar fidelity.

Several aspects of this work invite further investigation. The
symmetry properties of the discovered solution families warrant
systematic characterization: understanding why certain rotation
angles favor symmetric over antisymmetric solutions, and how
these families relate to the topology of the control landscape
may yield design principles for pulse optimization that can be
inaccessible to local gradient methods alone. 

A central advantage of this approach is its
adaptability. It requires no labeled training data — the system
Hamiltonian alone defines the optimization objective — and any
parameter entering the Hamiltonian or loss function can be
included as a network input. This makes the framework directly
transferable to other quantum platforms: replacing the
Hamiltonian and adjusting the network output is sufficient to
apply the method to a new physical system, with the training
procedure remaining unchanged. Extension to multi-qubit gates is
a natural next step, where the combinatorial growth of the
parameter space makes conventional pulse-by-pulse optimization
increasingly impractical, and the ability to generate pulses
across a continuous parameter space from a single trained network
becomes particularly valuable. A further direction is to couple this open-loop framework with closed-loop optimization: the network's pulses provide high-quality initial guesses that can then be refined on quantum hardware using feedback to adapt them to the specific imperfections of a given quantum device.

\section{Authors' Contributions}
T.K., F.M., and S.J.G. conceived the project together. T.K.
developed the neural network framework, performed network
training and benchmarking, and wrote the manuscript. T.H.
identified and analyzed the structural properties of
GRAPE-optimized pulses. L.V.D. developed an optimized
implementation of the GRAPE algorithm. S.H. contributed to
neural network optimization. F.M. and S.J.G. supervised the
project. All authors discussed the results and reviewed the
manuscript.
 \section{Acknowledgments}
The authors acknowledge funding from the Munich Quantum Valley (K-4, THEQUCO and K-8, HAT), which is supported by the Bavarian state government with funds from the Hightech Agenda Bayern Plus. Research was performed at the Bavarian NMR Center
(BNMRZ) at the Technical University of Munich.
\appendix
\section{Numerical Implementation}
\label{app:optimizations}

For a piecewise constant Hamiltonian over a time interval
$\Delta t$, we exploit the simpler structure of two-level
systems to avoid matrix exponentiation and matrix multiplication
entirely. A propagator is fully characterized by two
complex coefficients $(\alpha_n, \beta_n)$ satisfying
$|\alpha_n|^2 + |\beta_n|^2 = 1$ as $P_n=\begin{pmatrix}
    \alpha_n & -\beta^*_n \\
    \beta_n & \alpha^*_n
\end{pmatrix}$ with:
\begin{align}
    \alpha_n &= \cos\!\left(\frac{\omega_n \Delta t}{2}\right)
    - i\,\frac{\delta}{\omega_n}
    \sin\!\left(\frac{\omega_n \Delta t}{2}\right), \\
    \beta_n &= -i\,\frac{s\nu}{\omega_n}
    \sin\!\left(\frac{\omega_n \Delta t}{2}\right)\, e^{i\phi_n},
\end{align}
where $\omega_n = \sqrt{\delta^2 +
(s\nu)^2}$ is the effective Rabi frequency
and $\phi_n$ is the control phase at time step $n$. The
evolution operator is thus given by $U_n=\begin{pmatrix}
    A_n & -B^*_n \\ B_n & A_n^*
\end{pmatrix}$, where $A_n$ and $B_n$ satisfy the series
\begin{align}
    A_{n+1} &= \alpha_n\, A_n - \beta_n^*\, B_n, \\
    B_{n+1} &= \beta_n\, A_n + \alpha_n^*\, B_n,
\end{align}
with initial conditions $A_0 = 1$, $B_0 = 0$. This replaces all
$2 \times 2$ matrix operations with scalar complex arithmetic,
reducing each propagation step to four complex multiplications
and two additions. Since $\alpha_n$ and the phase-independent
part of $\beta_n$ depend only on the system parameters and
$\Delta t$, they are precomputed once before the time loop; at
each step, only the phase factor $e^{i\phi_n}$ is applied. The
final infidelity is obtained directly from the accumulated
coefficients $(A_N, B_N)$ and the target gate coefficients
$(A_T, B_T)$ without reconstructing any matrices.
\subsection{Exact Analytical Gradients}
The gradients given in Eq.~\eqref{eq:grape_gradient} require the evaluation of the derivative $\frac{\partial P_n}{\partial u_{kn}}$.
Standard GRAPE implementations approximate this derivative to first order in $\Delta t$ as:
\begin{equation}
    \frac{\partial \hat{P}_n}{\partial u_{kn}}
    = -i\,\Delta t\,\frac{\sigma_k}{2}\,\hat{P}_n+O(\Delta t^2).
\end{equation}
This first-order approximation requires small $\Delta t$ to
remain accurate. We
instead use the exact analytical derivative for the two-level
system. Defining $\frac{\partial \hat{P}_n}{\partial u_{kn}} =
\hat{A}_{kn}\,\hat{P}_n$, the matrices $\hat{A}_{kn}$ are given
by
\begin{equation}
\begin{split}
    \hat{A}_{kn} = -\frac{i}{2\omega_n} \Bigl[
    &\sin(\omega_n \Delta t)\,\sigma_k
    - [1 - \cos(\omega_n \Delta t)]\,\hat{M}_k \\
    &+ [\omega_n \Delta t - \sin(\omega_n \Delta t)]\,
    n_k\,(\vec{n} \cdot \vec{\sigma}) \Bigr],
\end{split}
\end{equation}
with cross-term matrices $\hat{M}_x = (n_y \sigma_z - n_z
\sigma_y)$ and $\hat{M}_y = (n_z \sigma_x - n_x \sigma_z)$.
Using these exact gradients allows arbitrarily large time steps $\Delta t$
without loss of accuracy.

In this work, we assume pulses of the form $u_x=\nu\cos\phi$ and $u_y=\nu\sin\phi$, where $\nu$ is kept constant and only the phase $\phi$ is optimized. The gradients $\partial \mathcal{J}/\partial\phi_n$ are calculated using the chain rule:
\[\begin{aligned}
\frac{\partial \mathcal{J}}{\partial\phi_n} & =\frac{\partial \mathcal{J}}{\partial u_{xn}}\frac{\partial u_{xn}}{\partial \phi_{n}}+\frac{\partial \mathcal{J}}{\partial u_{yn}}\frac{\partial u_{yn}}{\partial \phi_{n}}\\
& = -\frac{\partial \mathcal{J}}{\partial u_{xn}}\nu\sin\phi_n+\frac{\partial \mathcal{J}}{\partial u_{yn}}\nu\cos\phi_n.
\end{aligned}
\]

\section{Symmetry Operations on Optimized Pulses}
\label{app:symmetry}

Broadband universal rotation pulses can be transformed by symmetry operations
that leave the ensemble-averaged fidelity
invariant~\cite{Braun_2014}. The detuning ranges and rotation axes have direct implications on the allowed symmetry transformations. The set of valid symmetry operations depends on the target rotation angle $\beta$:

For all rotations with symmetric offset range and rotations about the y-axis, two symmetry operations are valid: time reversal, $\phi(t) \to \phi(T_\text{P} - t)$, and sign inversion of $u_x$, i.e.\ $u_x \to -u_x$. Both leave the ensemble-averaged fidelity unchanged~\cite{Braun_2014}.

For identity gates, i.e. $\beta = 2k\pi$ with $k \in \mathbb{Z}$, three additional symmetries arise: sign inversion of $u_y$, a global phase shift $\phi(t) \to \phi(t) + \phi_0$ for any constant $\phi_0$, and any cyclic permutation of the time dependent controls. None of these operations change the fidelity~\cite{Braun_2014}.

By systematically applying these operations to sequences of
pulses ordered by duration, one can reveal structural
continuities that are otherwise hidden by the arbitrary choices
made during independent GRAPE optimizations. We describe the
curation procedure for two representative cases:
all rotations which are not identity gates and $\beta=2\pi$.

Algorithm~\ref{alg:full_cleanup_arb_rotangle} and Algorithm~\ref{alg:full_cleanup_2pi} describe the curation procedures for rotation angles which are not identity gates and for $\beta=2\pi$, respectively. The curation procedure used in this work for $2\pi$ rotations exploits the symmetric pulse structure found in GRAPE pulses and the allowed symmetry transformations to align pulses across different durations into a visually continuous family. Since pulses of neighboring durations differ in length, direct
comparison requires interpolation: the control components
$(u_x, u_y)$ of shorter pulses are resampled to match the length
of the current pulse using cubic spline interpolation. The functions used in the algorithms are defined as follows:

\begin{enumerate}
\item \textbf{CyclicAlignment} For each
pulse, all possible cyclic shifts are evaluated and the shift
that minimizes the asymmetry
$\sum_t |u(t) - u(T_\text{P} - t)|^2$ is selected, where
$u(t) = u_x(t) + i\,u_y(t)$. This aligns each pulse to the most
symmetric orientation.

\item \textbf{TrendUX} For each
pulse, the two candidates $\varphi(t)$ and $\varphi(t, u_x \to -u_x)$ are compared against the preceding
pulses (using a lookback window). The candidate whose control
components are closest to the interpolated neighbors is retained.
This resolves the sign ambiguity in $u_x$ consistently across
the duration axis.

\item \textbf{TrendUY} The same procedure is applied to resolve the sign ambiguity in
$u_y$, comparing $\varphi(t)$ against
$\varphi(t,u_y \to -u_y)$.

\item \textbf{PhaseCorrection} A phase
shift is applied so that the phase at the pulse midpoint equals
$\pi/2$, establishing a common reference across all durations.

\item \textbf{TrendCircShift}  For each pulse, all combinations of cyclic shifts (multiples of $T_\text{P}/4$) and sign operations $(u_x, u_y) \to (\pm u_x, \pm u_y)$ are evaluated by comparing against interpolated neighboring pulses using a lookback window. The combination minimizing $\sum_t |u(t) - u_\text{prev}(t)|^2$ is selected by majority vote.

\item \textbf{TrendTimeReversal} For each pulse, the original and the time-reversed pulse are compared against interpolated neighboring pulses using the
lookback procedure. The version minimizing the distance to the neighbors is retained.
\end{enumerate}

Figure~\ref{fig: 2x2_cleaned_phase_profiles} demonstrates the
effect of these curation procedures. Panels~(a) and~(b) show the
uncurated and curated phase profiles for $\beta = 3\pi/2$,
panels~(c) and~(d) for $\beta = 2\pi$. In both cases, the
curation reveals continuous structural patterns across pulse
durations that are obscured in the raw GRAPE output by arbitrary
symmetry choices. The effect is particularly pronounced for
$2\pi$ rotations, where the additional cyclic symmetry produces
the highly fragmented appearance before curation.
 
\begin{algorithm}
\caption{Phase curation pipeline for arbitrary non-identity gates}
\label{alg:full_cleanup_arb_rotangle}
\begin{algorithmic}[1]
\Require $\boldsymbol{\Phi}$ (Array of phase profiles), $\mathbf{A}$ (Array of amplitude profiles), $L_\text{b}$ (Lookback window)
\Ensure $\boldsymbol{\Phi}_{\text{out}}$, $\mathbf{A}_{\text{out}}$ (Processed phase and amplitude profiles)
\State $\mathbf{L} \gets$ Count non-zero elements along temporal axis of $\boldsymbol{\Phi}$
\State $\boldsymbol{\Phi}_{\text{tr}} \gets \Call{TrendTimeReversal}{\boldsymbol{\Phi}, \mathbf{A}, \mathbf{L}, L_\text{b}}$
\State $\boldsymbol{\Phi}_{\text{out}} \gets \Call{TrendUX}{\boldsymbol{\Phi}_{\text{tr}}, \mathbf{A}, \mathbf{L}, L_\text{b}}$
\State $\mathbf{A}_{\text{out}} \gets \mathbf{A}$
\State \Return $\boldsymbol{\Phi}_{\text{out}}$, $\mathbf{A}_{\text{out}}$
\end{algorithmic}
\end{algorithm}

\begin{algorithm}
\caption{Phase curation procedure for $2\pi$ rotations}
\label{alg:full_cleanup_2pi}
\begin{algorithmic}[1]
\Require $\boldsymbol{\Phi}$ (Array of phase profiles), $\mathbf{A}$ (Array of amplitude profiles), $L_\text{b}$ (Lookback window)
\Ensure $\boldsymbol{\Phi}_{\text{out}}$, $\mathbf{A}_{\text{out}}$ (Processed phase and amplitude profiles)

\State $\mathbf{L} \gets$ Count non-zero elements along temporal axis of $\boldsymbol{\Phi}$ \Comment{Determine pulse lengths}

\State $\boldsymbol{\Phi}_{\text{shift}}, \mathbf{A}_{\text{shift}} \gets \Call{CyclicAlignment}{\boldsymbol{\Phi}, \mathbf{A}, \mathbf{L}}$

\State $\boldsymbol{\Phi}_{\text{ux-trend}} \gets \Call{TrendUX}{\boldsymbol{\Phi}_{\text{shift}}, \mathbf{A}_{\text{shift}}, \mathbf{L}}$

\State $\boldsymbol{\Phi}_{\text{uy-trend}} \gets \Call{TrendUY}{\boldsymbol{\Phi}_{\text{ux-trend}}, \mathbf{A}_{\text{shift}}, \mathbf{L}}$

\State $\boldsymbol{\Phi}_{\text{corr1}} \gets \Call{PhaseCorrection}{\boldsymbol{\Phi}_{\text{uy-trend}}, \mathbf{L}}$

\State $\boldsymbol{\Phi}_{\text{ux-trend2}} \gets \Call{TrendUX}{\boldsymbol{\Phi}_{\text{corr1}}, \mathbf{A}_{\text{shift}}, \mathbf{L}}$

\State $\boldsymbol{\Phi}_{\text{circ}}, \mathbf{A}_{\text{circ}} \gets \Call{TrendCircShift}{\boldsymbol{\Phi}_{\text{ux-trend2}}, \mathbf{A}_{\text{shift}}, \mathbf{L}, L_{\text{b}}=1}$

\State $\boldsymbol{\Phi}_{\text{corr2}} \gets \Call{PhaseCorrection}{\boldsymbol{\Phi}_{\text{circ}}, \mathbf{L}}$

\State $\boldsymbol{\Phi}_{\text{out}} \gets \Call{TrendUX}{\boldsymbol{\Phi}_{\text{corr2}}, \mathbf{A}_{\text{shift}}, \mathbf{L}}$ 

\State $\mathbf{A}_{\text{out}} \gets \mathbf{A}_{\text{circ}}$

\State \Return $\boldsymbol{\Phi}_{\text{out}}$, $\mathbf{A}_{\text{out}}$
\end{algorithmic}
\end{algorithm}

\section{Model Size Analysis}
\label{app:model_size}

To assess how strongly our results depend on the network
architecture, we trained networks of varying depth ($5$ or $7$
hidden layers) and width ($512$, $1024$, or $2048$ neurons per
layer) under otherwise identical conditions. As in the main
text, each architecture consists of two networks covering the
lower ($[0,\pi]$) and upper ($(\pi,2\pi]$) halves of the
rotation-angle range and trained with the same parameters.
All variants were evaluated on the same set of $100{,}000$
randomly sampled gate configurations from $\mathcal{G}$, with
the infidelity of each configuration averaged over a grid of
offset and amplitude deviations. Table~\ref{tab:model_size}
reports the absolute numbers of test configurations reaching an
infidelity below $10^{-2}$, $10^{-3}$, and $10^{-4}$. The
performance is largely insensitive to the architecture with the model with 5x2048 being the overall best choice.

\begin{table}
\caption{Comparison of network architectures. All networks were
trained under identical conditions and evaluated on the same $100{,}000$ randomly sampled test
configurations. Values give the total number of configurations
whose infidelity $\mathcal{J}$ falls below the indicated
threshold.}
\label{tab:model_size}
\begin{ruledtabular}
\begin{tabular}{lcccc}
Architecture & Parameters & ${\mathcal{J}}< 10^{-2}$  & ${\mathcal{J}}< 10^{-3}$ & ${\mathcal{J}}< 10^{-4}$  \\
\hline
5 $\times$ 512  & $\sim$1.5M & 18{,}329 & 8{,}429 & 3{,}896 \\
7 $\times$ 512  & $\sim$2.0M & 18{,}110 & 8{,}212 & 3{,}664  \\
5 $\times$ 1024 & $\sim$5.1M & 18{,}196 & 8{,}577 & 4{,}103 \\
7 $\times$ 1024 & $\sim$7.2M & 18{,}231 & 8{,}507 & 4{,}011  \\
5 $\times$ 2048 & $\sim$18.6M & 18{,}438 & 8{,}655 & 4{,}069  \\
7 $\times$ 2048 & $\sim$27.0M  & 18{,}366 & 8{,}617 & 3{,}961\\
\end{tabular}
\end{ruledtabular}
\end{table}

\section{Computational Cost Analysis}
\label{app:computation}
To provide a transparent comparison of computational costs, we
summarize the resources required by each method and discuss the
conditions under which the neural network approach amortizes its
upfront training investment. We note that precise quantum control
simulations require double-precision (FP64) arithmetic to
maintain numerical accuracy. Consumer GPUs such as the RTX 5090 deliver only
$\sim$1.6\,TFLOPS of FP64 throughput (trillions of
floating-point operations per second), compared to
$\sim$105\,TFLOPS in single precision (FP32) --- a 64-fold
reduction. This makes consumer GPUs poorly suited for
double-precision workloads such as GRAPE or NN training with FP64. The dual-socket AMD
EPYC 9554 high-performance server node, providing $\sim$6.3\,TFLOPS of FP64
throughput, therefore represents the natural hardware choice for
GRAPE.
The key distinction between the two approaches is that GRAPE's
cost scales linearly with the number of pulses, whereas the
network's cost is dominated by a one-time training investment.
Once trained, generating a pulse requires only milliseconds of GPU time and seconds for 100{,}000 pulses. The training cost of
5 days and 1 hour GPU time therefore amortizes fast with interpolation in the parameter space: This becomes particularly advantageous in
scenarios requiring repeated pulse generation, such as
recalibration of experimental parameters, exploration of
different robustness specifications, or deployment across
multiple devices with varying hardware characteristics. We emphasize that this is not a hardware-controlled comparison
--- each method runs on the platform best suited to its
computational structure. A direct GPU-vs-GPU or CPU-vs-CPU
comparison would disadvantage one method without providing
additional insight, since GRAPE does not benefit substantially
from GPU acceleration (each pulse optimization is sequential)
and the neural network does not benefit from many-core CPU
parallelism during inference.

\bibliography{quellen}

\end{document}